\documentclass[showpacs,preprintnumbers,amsmath,amssymb,twocolumn]{revtex4}

\usepackage{graphicx}
\usepackage{dcolumn}
\usepackage{bm}

\begin{document}

\title{Sound Velocity and Meissner Effect in Light-heavy Fermion Pairing Systems}
\author{Lianyi He, Meng Jin, and Pengfei Zhuang\\
        Physics Department, Tsinghua University, Beijing 100084, China}

\begin{abstract}
In the frame of a four fermion interaction theory, we investigated
the collective excitation in light-heavy fermion pairing systems.
When the two species of fermions possess different masses and
chemical potentials but keep the same Fermi surface, we found that
the sound velocity in superfluids and the inverse penetration
depth in superconductors have the same mass ratio dependence as
the ratio of the transition temperature to the zero temperature
gap.
\end{abstract}
\maketitle

The study on Cooper pairing between fermions with different masses
promoted great interest both theoretically and experimentally in
recent years. A new pairing phenomenon, the breached pairing or
interior gap, was proposed in the study of light-heavy fermion
pairing systems\cite{BP1,BP2,BP3}. The light-heavy fermion pairing
can exist in an electron system where the electrons are from
different bands, a cold fermionic atom gas where a mixture of
$^6$Li and $^{40}$K can be realized, and a color superconductor
with strange quarks.

In this Letter, we focus on the Cooper pairing between two species
of fermions with equal Fermi surfaces but unequal masses. Such
constraint on the fermion pairing can be realized, for instance,
in the mixed atom gas of $^6$Li and $^{40}$K by adjusting the
numbers of the two species to be the same. It is recently found
that, for such a system the ratio between the transition
temperature $T_c$ and the zero temperature gap $\Delta_0$ depends
only on the mass ratio $\alpha=m_b/m_a$ between the two masses
$m_a$ and $m_b$\cite{caldas},
\begin{equation}
\label{alpha1}
{T_c\over \Delta_0}={e^\gamma\over
\pi}\left({2\sqrt{\alpha}\over 1+\alpha}\right)
\end{equation}
with the Euler constant $\gamma$. Without loss of generality, we
assume $m_b > m_a$ in the following.

What is the behavior of the collective excitation in such a
system, and does the simple $\alpha$ dependence still hold when we
go beyond the mean field approximation? In a superfluid composed
of neutral fermions, the low energy excitation is the Goldstone
mode which is directly related to the specific heat at low
temperature. In a superconductor composed of charged fermions, the
collective mode is associated with the Meissner effect.

We consider a system composed of two species of fermions with
attractive interaction, described by the Lagrangian density in
Euclidean space $x=(\tau=it,{\bf x})$,
\begin{equation}
\label{lag} {\cal L}=\sum_{i=a,b}\psi_{i}^*\left(-\partial_
\tau+\frac{\nabla^2}{2m_i}+\mu_i\right)\psi_{i}+g\psi_{a}^*\psi_{b}^*\psi_{b}\psi_{a},
\end{equation}
where $\psi_i(x)$ describe the fermion fields, $g$ is the coupling
constant, and $\mu_a$ and $\mu_b$ are the chemical potentials of
the two species. We will take the units $c=\hbar =k_B=1$
throughout the paper.

We can perform an exact Hubbard-Stratonovich transformation to
introduce an auxiliary boson field $\phi(x)$ and its complex
conjugate $\phi^*(x)$. By defining fermion field vector $
\Psi^*=\left(\begin{array}{cc} \psi_a^*, &
\psi_b\end{array}\right)$ in the Nambu-Gorkov space, the partition
function of the system can be written as
\begin{equation}
Z=\int[d\Psi^*][d\Psi][d\phi][d\phi^*]e^{\int_0^\beta d\tau\int
d^3{\bf x}\left(\Psi^*K\Psi-{|\phi|^2\over g}\right)}
\end{equation}
with the kernel $K$ defined as
\begin{equation}
K[\phi]=\left(\begin{array}{cc} -\partial_
\tau+\frac{\nabla^2}{2m_a}+\mu_a&\phi(x)
\\ \phi^*(x)&-\partial_\tau-\frac{\nabla^2}{2m_b}-\mu_b\end{array}\right),
\end{equation}
where $\beta$ is the inverse of the temperature, $\beta=1/T$.

When the interaction is turned off, the chemical potentials
$\mu_a$ and $\mu_b$ are regarded as the corresponding Fermi
energies. Since we focus on the Cooper pairing with equal Fermi
surfaces, the Fermi momenta of the two species are the same,
$p_F^a=p_F^b=p_F=\sqrt{2m_a\mu_a}=\sqrt{2m_b\mu_b}$, and the
number densities are also the same, $n_a=n_b=n/2=p_F^3/(6\pi^2)$.
In this case, the breached pairing state which is unstable to
phase separation \cite{bedaque} and LOFF state\cite{LOFF} is ruled
out.

For the zero range interaction in (\ref{lag}), we need a
regularization scheme. For a solid, we can suppose that the
attractive interaction is restricted in a narrow momentum region
around the Fermi surface, $p_F-\Lambda<|{\bf p}|<p_F+\Lambda$ with
$\Lambda\ll p_F$, like the classical BCS theory and the study in
\cite{BP1} where $\Lambda$ serves as a natural ultraviolet cutoff
in the theory. For a dilute fermionic atom gas, we can replace the
bare coupling $g$ by the low energy limit of the two-body
scattering matrix\cite{BP3}, namely
\begin{equation}
\frac{m}{4\pi a_s}=-\frac{1}{g}+\int{d^3{\bf p}\over
(2\pi)^3}\frac{1}{2\epsilon_p}
\end{equation}
with the s-wave scattering length $a_s$, $\epsilon_p=p^2/(2m)$,
and the reduced mass $m=2m_am_b/(m_a+m_b)$. While the results from
the regularization scheme with the cutoff $\Lambda$ and the scheme
with the scattering length $a_s$ are the same in weak interaction
limit, the latter is also a good approach to study the BCS-BEC
crossover where the interaction is strong. Without loss of
generality, we adopt the latter in this Letter.

\begin{figure}
\centering
\includegraphics[width=5cm]{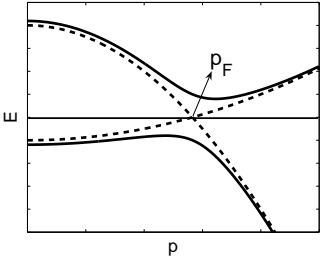}
\caption{A schematic description of the dispersion relations for
the two branches of free fermions (dashed lines) and
quasiparticles (solid lines). $p_F$ is the common Fermi momentum.
\label{fig1}}
\end{figure}

In mean field approximation, the boson field $\phi$ is replaced by
its vacuum expectation value $\Delta$ which can be chosen to be
real, and the gap equation which determines $\Delta$ is derived
from the minimum of the thermodynamical potential $\Omega=-T\ln
Z$. At zero temperature, it reads
\begin{equation}
\label{gap1}
\frac{-m}{4\pi a_s}=\int_0^\infty
dp\frac{p^2}{2\pi^2}\left(\frac{1}{2E_p}-\frac{1}{2\epsilon_p}\right),
\end{equation}
where $E_p$ is defined as $E_p=\sqrt{(\epsilon_p-\mu)^2+\Delta^2}$
with the average chemical potential $\mu=(\mu_a+\mu_b)/2$. In the
weak coupling limit, the solution of the gap equation at zero
temperature can be analytically expressed as
\begin{equation}
\Delta_0=8e^{-2}\frac{p_F^2}{2m} e^{\frac{\pi}{2p_Fa_s}}.
\end{equation}
The fermion propagator ${\cal G}\equiv K^{-1}$ in the Nambu-Gorkov
space can be explicitly written as
\begin{equation}
{\cal G}(i\nu_n,{\bf
p})={(i\nu_n-\xi_p^-)+\tau_3\xi_p^+-\tau_1\Delta\over
(i\nu_n-\xi_p^-)^2-E_p^2},
\end{equation}
where $\nu_n$ is the fermion Matsubara frequency, $\tau_i$ are the
Pauli matrices, and $\xi_p^\pm$ are defined as
\begin{equation}
\xi_{p}^\pm=\frac{1}{2}(\xi_p^a\pm\xi_p^b)
\end{equation}
with the free fermion dispersion relations
$\xi^a_p=p^2/(2m_a)-\mu_a$ and $\xi^b_p=p^2/(2m_b)-\mu_b$. The
dispersion relations of the quasiparticles can be read from the
poles of the propagator. A schematic description for the two
branches of quasiparticles is illustrated in Fig.\ref{fig1}. Due
to the constraint of equal Fermi surfaces on the pairing, all
fermionic excitations are gapped, as in the standard BCS theory
with equal fermion masses.

The transition temperature $T_c$ is determined by the gap equation
at finite temperature and at $\Delta=0$,
\begin{equation}
\label{gap2}
\frac{-m}{4\pi a_s}=\int_0^\infty
dp\frac{p^2}{2\pi^2}\left(\frac{1-f(\xi_p^a)-f(\xi_p^b)}{2(\epsilon_p-\mu)}-\frac{1}{2\epsilon_p}\right),
\end{equation}
where $f(x)=1/\left(e^{x/T}+1\right)$ is the Fermi-Dirac
distribution function. In the case of weak coupling, the number
equations can be approximated by the condition $m_a\mu_a=m_b\mu_b$
which gives $\xi_p^a=2(\epsilon_p-\mu)/(1+\alpha)$ and
$\xi_p^b=\alpha\xi_p^a$. With the standard trick\cite{fetter}, we
can reobtain the $\alpha$-dependence of the ratio $T_c/\Delta_0$
as shown in Eq. (\ref{alpha1}). For $\alpha=1$, we recover the
standard BCS result $T_c\simeq0.57\Delta_0$. If $n$ and $a_s$ are
fixed, we have the relation between the two critical temperatures,
\begin{equation}
T_c(m_a,m_b)=\sqrt{m_a/m_b}T_c(m_a,m_a),
\end{equation}
which means that the critical temperature for the mixed $^6$Li and
$^{40}$K system is about $\sqrt{1/7}$ of the one for the pure
$^6$Li system.

We now start to investigate the low energy collective excitation
in the superfluid state. The spontaneous symmetry breaking and the
associated Goldstone mode in the superfluid state provides an
effective field theory approach for the collective
excitation\cite{weinberg,nao,liu}. In our case, the particle
number is conserved which corresponds to a global $U(1)$ symmetry
of the phase transformation,
\begin{equation}
\psi_{i}(x)\rightarrow e^{i\varphi}\psi_{i}(x), \ \ \
\phi(x)\rightarrow e^{2i\varphi}\phi(x)
\end{equation}
with an arbitrary and constant phase $\varphi$. The nonzero
condensate $\Delta$ of Cooper pairs in the superfluid state
spontaneously breaks the $U(1)$ symmetry, and correspondingly, a
Goldstone mode which possesses linear dispersion at low energy is
expected to appear. The low energy dynamics of the system at low
temperature is then dominated by the Goldstone mode. Since all
fermions are gapped in our case, the fluctuation of the amplitude
of the order parameter at low temperature can be neglected.
Therefore, we can write the order parameter field as
\begin{equation}
\phi(x)=\Delta e^{2i\theta(x)}.
\end{equation}
Taking the standard approach\cite{nao,liu}, we can transform
nonperturbatively the fermion fields as follows,
\begin{equation}
\psi_{i}(x)=\tilde{\psi}_{i}(x)e^{i\theta(x)}, \ \ \
\psi_{i}^*(x)=\tilde{\psi}_{i}^*(x)e^{-i\theta(x)}.
\end{equation}
The transformation is designed to eliminate the phase fluctuation
$\theta(x)$ dependence from the off-diagonal pairing potential
terms $(\psi_{a}^*\psi_{b}^*\phi+c.c.)$ in the Lagrangian. The
$\theta(x)$ dependence can appear only in the kinematic terms of
the fermion sector, and the Lagrangian density of the system can
be expressed as
\begin{equation}
{\cal
L}=\tilde{\psi}_{i}^*\left(-D_{\tau}+\frac{\vec{D}^2}{2m_i}-\mu_i\right)\tilde{\psi}_{i}
+(\Delta\tilde{\psi}_{b}\tilde{\psi}_{a}+c.c.)+\frac{\Delta^2}{g}
\end{equation}
with $D_{\mu}=\partial_\mu+i\partial_\mu\theta(x)$. Using the
Nambu-Gorkov vector, the partition function can be rewritten as
\begin{equation}
Z=\int[d\Psi^*][d\Psi][d\theta]e^{\int_0^\beta d\tau\int d^3{\vec
x}\left(\Psi^*{\cal K}[\theta]\Psi+{\Delta^2\over g}\right)}
\end{equation}
with the kernel ${\cal K}[\theta]$ defined as
\begin{equation}
{\cal
K}[\theta]=K[\Delta]-\tau_3i\partial_\tau\theta-\Sigma_-(\nabla\theta)^2+\Sigma_+i\nabla\theta\cdot\nabla,
\end{equation}
where the matrices $\Sigma_\pm$ are defined as
\begin{equation}
\Sigma_\pm=\frac{1}{2}\left(\begin{array}{cc}1/m_a & 0\\
0 & \pm1/m_b \end{array}\right)
\end{equation}

Since the fermionic excitations are gapped at all energies below
$\Delta$, for collective excitation with energy below $2\Delta$,
we can safely integrate out all fermionic degrees of freedom and
obtain an effective action for the collective mode only
\begin{equation}
S[\theta]=\int_0^\beta d\tau\int d^3{\bf x}\ \textrm{Tr}\ln {\cal
K}[\theta],
\end{equation}
where the trace is taken over the fermion momentum, frequency and
Nambu-Gorkov vector. We have neglected here a constant which is
irrelevant for the following discussions.

The next task is to expand the action in powers of $\theta(x)$. We
use the standard derivative expansion. With the two matrices $V_1$
and $V_2$ defined in the fermion momentum, frequency and
Nambu-Gorkov space\cite{nao},
\begin{eqnarray}
(V_1)_{k,k^\prime}&=&\frac{1}{\sqrt{\beta
V}}(\nu_n-\nu_{n^\prime})\theta(k-k^\prime)\tau_3\nonumber\\
&+&\frac{1}{\sqrt{\beta
V}}i({\bf k}+{\bf k^\prime})\cdot({\bf k}-{\bf k^\prime})\theta(k-k^\prime)\Sigma_+\nonumber\\
(V_2)_{k,k^\prime}&=&\frac{1}{2\beta
V}\sum_{k^{\prime\prime}}({\bf k}-{\bf k^{\prime\prime}})\cdot({\bf k^{\prime\prime}}-{\bf k^{\prime}})\nonumber\\
&\times&\theta(k-k^{\prime\prime})\theta(k^{\prime\prime}-k^{\prime})\Sigma_-,
\end{eqnarray}
we can expand the effective action to any order of $\theta$. Since
we consider only the low energy behavior of the collective mode,
we need only the expansion up to the quadratic term,
\begin{equation}
S[\theta]=\textrm {Tr}[{\cal G}V_2]+\frac{1}{2}\textrm {Tr}[{\cal
G}V_1{\cal G}V_1].
\end{equation}
After a straightforward calculation, we have
\begin{equation}
S[\theta]=\sum_q\left[-\frac{{\cal D}(q)}{2}\omega_n^2+\left({\cal
P}(q)+\eta\right)\frac{{\bf q}^2}{2}\right]|\theta(q)|^2
\end{equation}
with the four momentum $q=(i\omega_n,{\bf q})$ of the collective
mode, $\eta=n_a/m_a+n_b/m_b$, and
\begin{eqnarray}
{\cal D}(q)&=&\frac{1}{\beta V}\sum_p\textrm {Tr}\left[\tau_3{\cal
G}(p)\tau_3{\cal G}(p+q)\right],\\
{\cal P}(q)&=&\frac{1}{\beta V}\sum_p\frac{{\bf p}^2}{3}\textrm
{Tr}\left[\Sigma_+{\cal G}\left(p-\frac{q}{2}\right)\Sigma_+{\cal
G}\left(p+\frac{q}{2}\right)\right].\nonumber
\end{eqnarray}
Now the trace is taken only over the Nambu-Gorkov vector. We
should note that the functions ${\cal D}$ and ${\cal P}$ are
related to the density and current correlation
functions\cite{nao,liu} $\langle\rho(x)\rho(0)\rangle$ and
$\langle J_i(x)J_j(0)\rangle$, respectively. Also, we want to
emphasize that due to the mass difference, the function ${\cal P}$
is quite different from its usual form in the case with equal
masses.

We now discuss the low energy behavior of the collective mode. The
energy of the collective mode is defined through the analytical
continuation $i\omega_n\rightarrow \omega+i0^+$. In the low energy
limit, namely $\omega\ll 2\Delta$ and $v_F|{\bf p}|\ll 2\Delta$,
where the Fermi velocity $v_F$ is defined as $v_F=p_F/m$, we can
expand the functions ${\cal D}$ and ${\cal P}$ in powers of the
momentum ${\bf q}$. To the leading order, the effective action
becomes
\begin{eqnarray}
S[\theta]&=&\frac{1}{2}\sum_q\left(A\omega_n^2+B{\bf
q}^2\right)|\theta(q)|^2,\nonumber\\
A&=&\int_0^\infty dp\frac{p^2}{2\pi^2}\frac{\Delta^2}{E_p^3},\nonumber\\
B&=&\frac{n}{m}-\frac{1}{3}\left(\alpha-1\over
\alpha+1\right)^2\int_0^\infty
dp\frac{p^4}{2\pi^2m^2}\frac{\Delta^2}{E_p^3}.
\end{eqnarray}
Note that the second term of $B$ comes from ${\cal P}(0)$ which
vanishes for symmetric systems with $\alpha=1$. The energy
spectrum of the collective mode is obtained by finding the zero of
the coefficient of the quadratic term in $\theta$,
\begin{equation}
\omega=v_s|{\bf q}|
\end{equation}
with the definition for the sound velocity $v_s=\sqrt{B/A}$. In
weak coupling limit, the chemical potentials $\mu_a$ and $\mu_b$
of the two species are approximately their Fermi energies, and the
average chemical potential $\mu$ is much larger than the gap
parameter, the integrations in $A$ and $B$ can then be integrated
out,
\begin{equation}
A\simeq\frac{mp_F}{\pi^2},\ \ \
B\simeq\frac{n}{m}-\frac{1}{3}\left(\alpha-1\over
\alpha+1\right)^2\frac{p_F^3}{\pi^2m}.
\end{equation}
Together with the total fermion number $n=p_F^3/(3\pi^2)$, the
sound velocity has the same mass ratio dependence as the ratio
$T_c/\Delta_0$,
\begin{equation}
\label{alpha2}
v_s={v_F\over\sqrt 3}\left({2\sqrt\alpha\over
1+\alpha}\right).
\end{equation}
In the limit $\alpha\rightarrow 1$, we recover the well known
result $v_s=v_F/\sqrt{3}$. In the other limit $\alpha\rightarrow
\infty$, the sound velocity tends to be zero.

At sufficient low temperature $T\ll T_c$, the specific heat $c$ is
dominated by the Goldstone mode which has the $T^3$ power law
\begin{eqnarray}
c=\frac{2\pi^2}{15v_s^3}T^3,
\end{eqnarray}
which means that the specific heat for the mixed $^6$Li and
$^{40}$K system is about 17 times the one for the pure $^6$Li
system if $n$ is keep fixed. By experimentally measuring the
specific heat, we can determine the sound velocity and check the
theoretic result.

If the fermions carry electric charges, the Goldstone mode will
disappear due to the long range electromagnetic interaction
between the fermions\cite{nao}. In the language of gauge field
theory, the Goldstone mode is eaten up by the electromagnetic
field via Anderson-Higgs mechanism. This phenomenon is also called
Meissner effect. Suppose the two species carry charges $eQ_a$ and
$eQ_b$, respectively, the London penetration depth $\lambda_L$ is
related to the function ${\cal P}$\cite{he},
\begin{equation}
\frac{1}{4\pi\lambda_L^2}=\frac{ne^2}{2m}(Q_a^2+Q_b^2)+e^2{\cal
P}(0)
\end{equation}
with the modified matrix
\begin{equation}
\Sigma_+=\left(\begin{array}{cc}Q_a/m_a & 0\\
0 & Q_b/m_b \end{array}\right)
\end{equation}
appeared in ${\cal P}$ due to the charge difference between the
two species.

For a neutral Cooper pair with $Q_a=-Q_b=1$, the inverse of the
London penetration depth squared is zero as we expected,
\begin{equation}
\frac{1}{4\pi\lambda_L^2}=\frac{ne^2}{m}-\frac{e^2}{3}\int_0^\infty
dp {p^4\over 2\pi^2m^2}{\Delta^2\over E_p^3}=0.
\end{equation}
For the case with $Q_a=Q_b=1$, the mass ratio dependence is again
the same as the ratio $T_c/\Delta_0$ and the sound velocity $v_s$,
\begin{equation}
\label{alpha4} \lambda_L^{-1}=\sqrt{4\pi ne^2\over
m}\left({2\sqrt\alpha\over 1+\alpha}\right).
\end{equation}
In the limit $\alpha\rightarrow 1$, we recover the familiar
penetration depth $\lambda_L^{-1}=\left(4\pi ne^2/m\right)^{1/2}$.
On the other hand, in the limit $\alpha\rightarrow \infty$, the
penetration depth approaches infinity, which means an ideal
type-II superconductor. Since the penetration depth depends
strongly on the mass difference, it may change the type of
superconductors\cite{type}.
\begin{figure}
\centering
\includegraphics[width=5cm]{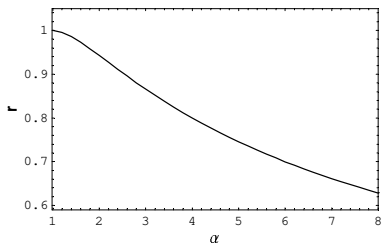}
\caption{The common $\alpha$ dependence of the ratio of the
transition temperature to the zero temperature gap, the sound
velocity, and the inverse penetration depth. \label{fig2}}
\end{figure}

For systems with fixed reduced mass $m$ (and hence fixed $v_F$)
but different mass ratio $\alpha$, from the equations
(\ref{alpha1}), (\ref{alpha2}) and (\ref{alpha4}), we can define a
quantity $r$ which describes the common mass ratio dependence of
$T_c/\Delta_0, v_s$ and $\lambda_L^{-1}$,
\begin{equation}
r={T_c(\alpha)/\Delta_0(\alpha)\over
T_c(1)/\Delta_0(1)}={v_s(\alpha)\over
v_s(1)}={\lambda_L^{-1}(\alpha)\over
\lambda_L^{-1}(1)}={2\sqrt\alpha\over 1+\alpha},
\end{equation}
and  plot it as a function of $\alpha$ in Fig.\ref{fig2}, where
$T_c(1), \Delta_0(1), v_s(1)$ and $\lambda_L(1)$ are quantities
for the symmetric system with equal masses $m_a=m_b=m$ and equal
chemical potentials $\mu_a=\mu_b=\mu$.

Based on a general four fermion interaction model, we have derived
the sound velocity in a superfluid and the penetration depth in a
superconductor for systems where the two species of fermions can
possess different masses and chemical potentials but keep the same
Fermi surface. We found that the sound velocity and the inverse
penetration depth have the same mass ratio dependence as the ratio
of the transition temperature to the zero temperature gap. While
our result is obtained in weak coupling BCS region, we expect that
the qualitative effect will still work in strong coupling systems,
such as the mixed atom gas of $^6$Li and $^{40}$K. In this Letter,
we did not consider the Fermi surface mismatch which can induce
exotic states such as breached pairing and LOFF states. In these
states, there exist gapless fermionic excitations and one can not
safely integrate out all fermionic degrees of freedom. When we
apply the above effective theory to such gapless states, we will
get imaginary sound velocity and penetration depth\cite{he}, which
have been widely discussed in the study of color
superconductivity\cite{huang,ren}.

{\bf Acknowledgments:}\ We thank Dr.H.Ren for helpful discussions
during the work. The work was supported by the grants
NSFC10425810, 10435080 and 10575058.

\end{document}